\documentclass[12pt]{article}
\usepackage{newlfont}
\usepackage{amsmath}
\usepackage{times}
\usepackage[all]{xypic}
\usepackage{array,multirow,graphicx}
\usepackage[colorlinks]{hyperref}
\usepackage{authblk}
\newcommand{\be}{\begin{equation}
\addtolength{\abovedisplayskip}{\extraspaces}
\addtolength{\belowdisplayskip}{\extraspaces}
\addtolength{\abovedisplayshortskip}{\extraspace}
\addtolength{\belowdisplayshortskip}{\extraspace}}
\newcommand{\ee}{\end{equation}}

\newcommand{\ba}{\begin{eqnarray}
\addtolength{\abovedisplayskip}{\extraspaces}
\addtolength{\belowdisplayskip}{\extraspaces}
\addtolength{\abovedisplayshortskip}{\extraspace}
\addtolength{\belowdisplayshortskip}{\extraspace}}
\newcommand{\ea}{\end{eqnarray}}

\numberwithin{equation}{section}
\begin{document}
\title{FRW in quadratic form of $f(T)$ gravitational theories}%
\author[1,2,3]{G.L. Nashed \thanks{nashed@bue.edu.eg}}
\affil[1]{\small \it Centre for Theoretical Physics, The British
University in Egypt, El-Sherouk City, Misr - Ismalia Desert Road,
Postal No. 11837, P.O. Box 43, Egypt.}
\affil[2]{\small \it Mathematics Department, Faculty of Science, Ain Shams University, Cairo, 11566, Egypt.}
\affil[3]{\small \it Egyptian Relativity Group, Egypt.}
\renewcommand\Authands{ and }
\maketitle
\begin{abstract}
We derive asymptote  solution of a homogeneous and isotropic universe governed by the quadratic form of the field equation of $f(T)$ gravity. We explain  how the higher order of the torsion  can provide an origin for  late accelerated phase  of the universe in the FRW.  The solution makes the scalar torsion  $T$ to be a  function of the cosmic  time $t$. We show that for the equation of state $p=\omega \rho$ with $\omega\neq -1$ the scale factor represent late phase of universe. We perform the cosmological studies and show how the quadratic form of $f(T)$ effect on the behavior of these studies.
\end{abstract}
\smallskip

{\bf keywords}: \textit{late phase of universe,  teleparallel gravity, accelerating universe, dark energy}.

\section{Introduction}\label{S1}

The method of describing the gravitational interaction can be done in different ways, the main and
best known is the Riemannian geometry as a tool for the formulation of this description.
This is known as the theory of general relativity (GR) of Einstein \cite{Wr1}.

 Riemannian geometry is based on the differential geometry which deals with the so-called
space-time as a differentiable manifold of dimension four, where one has  the effect of the
curvature and torsionless. There are other geometries that generalize this idea. A space-time may have curvature,
but also torsion in their structure. There are two fundamental concepts in differential geometry.
The formulation of this type of geometry has been undertaken and gravitation is known as Einstein-Cartan geometry \cite{SPT}. In this theory the gravitational interaction is described by both curvature and
torsion of space-time, where the torsion is commonly attributed to the inclusion of spin, through
fractional spin fields. A very particular case of this theory is when we take the identically zero
curvature, and then only have a space-time with torsion. This type of geometry is known as the
Weitzenb\"ock geometry \cite{To, Wr} where the torsion describes the gravitational interaction. Various
analysis can be performed in this type of geometry, which is proven to be dynamically equivalent
to GR  \cite{Mj}.

One extends gravity theory beyond general relativity is teleparallel gravity (TG). The birth of this gravity theory
refers back to 1928 \cite{Ea2}. At that time Einstein was trying to redefine the unification of gravity and electromagnetism
by introducing the notion of tetrad (vierbin) field together with the assumption of absolute parallelism. In this theory
the metric $g_{\mu \nu}$ is not the dynamical object, but one has a set of tetrad fields ${e^a}_\mu$, and instead of the well-known
torsionless Levi-Civita connection of GR theory, one works with a Weitzenb\"ok connection to introduce the covariant
derivative \cite{Wr}. Furthermore, the role of curvature scalar in GR is played by torsion scalar $T$ in the TG.

Recently, in order to account for the observed late time cosmic acceleration, the teleparallel Lagrangian density
described by the torsion scalar $T$ has been promoted to a function of $f(T)$ \cite{FF7}--\cite{CDDS}. Indeed, $f(T)$ theories can be used to
explain the cosmic acceleration and observations on large scales (e.g. via galaxy clustering
and cosmic shear measurements \cite{CCR}), but we must remember that since GR is in excellent
agreement with solar system and binary pulsar observations \cite{Wc}, every theory that aims
at explaining the large scale dynamics of the universe should reproduce GR in a suitable
weak-field limit: the same holds true for $f(T)$ theories. Recently, solar system data \cite{IS, XD}
have been used to constrain $f(T)$ theories; these results are based on the spherical symmetry
solution found in \cite{IS}, who used a diagonal tetrad. In this paper we follow
the approach described in \cite{TB} to define a ``good tetrad'' in $f(T)$ gravity - that is consistent
with the equations of motion without constraining the functional form of the Lagrangian -
and solve the field equations to obtain weak-field solutions with a power-law ansatz for an
additive term to the TEGR Lagrangian, $f(T) = T +\epsilon T^2$.

The main target of this work is to show how the quadratic form of $f(T)$ gravity can be useful in explaining the flatness and acceleration at  late phase of our universe. For as is well known, current observations of the present universe indicates that our universe now is almost spatially flat. This leads to exclude the closed and open universe models. On the other hand the initial flat space assumption contradicts the presence of the strong gravitational field (i.e. the Riemann curvature) as it should be! This contradiction might be explained as the flatness problem of the standard cosmology. Actually this problem has been overcome by the idea of an inflationary scenario during $\sim 10^{-36}-10^{-32}$ sec from the big bang. Lots of inflationary models have been proposed by using scalar fields. But to gain the benefits of both inflation and the standard cosmology the inflation should end at $\sim 10^{-32}$ sec from the big bang. This needs slow-roll conditions so that the inflationary universe ends with a vacuum dominant epoch allowing universe to restore the big bang scenario. So the inflation can be considered as an add-on tool rather a replacement of the big bang \cite{Le}. Until now there are no satisfactory reasons for the transition from inflation to big bang. Our trail here to treat these problems starts by diagnosing the core of the problem. We found that the curvature within the framework of the GR may lead to these conflicts, while introducing new qualities to the space-time, as torsion, might give a different insight of these problems.

The work is arranged as follows: In Section \ref{S2}, we describe the fundamentals of $f(T)$ gravity theory. We next show the contribution of the torsion scalar field to the density and the pressure of the FRW models and necessary modifications in Section \ref{S3}. Also we studied the case of quadratic form of  $f(T) = T +\epsilon T^2$ for many kinds of fluids for flat spacetime in Section \ref{S3}. We also investigate the cosmological behavior of the flat in  Section \ref{S3}. Moreover, we give the physical descriptions for the obtained results. In the flat universe and for the dark energy $\omega=-1$, the teleparallel torsion scalar field $T$ and the $f(T)$ appear as constant functions and the later might replace the cosmological constant, the universe shows an inflationary behavior as the scale factor $R(t) \propto e^{Ht}$, where the Hubble parameter $H$ is a constant. For $\omega\neq -1$, we get a scale factor that describes late universe in Section  \ref{S3}. In section  \ref{S4}, we discuss the cosmological parameter for the flat case  when $\omega \neq -1$.  Section \ref{S5} is devoted for summarizing and concluding the results.
\section{ABC  of $f(T)$}\label{S2}

In the Weitzenb\"{o}ck space-time the fundamental field variables describing gravity are a quadruplet of parallel vector fields \cite{Wr} ${h_i}^\mu$, which we call the tetrad field characterized by
\begin{equation}\label{q1}
D_{\nu} {h_i}^\mu=\partial_{\nu}
{h_i}^\mu+{\Gamma^\mu}_{\lambda \nu} {h_i}^\lambda=0,
\end{equation}
where
${\Gamma^\mu}_{\lambda \nu}$ define the nonsymmetric affine connection
\begin{equation}\label{q2}
{\Gamma^\lambda}_{\mu \nu} \stackrel{\textmd{def.}}{=}
{h_i}^\lambda {h^i}_{\mu, \nu},
\end{equation}
with $h_{i \mu, \nu}=\partial_\nu h_{i \mu}$\footnote{space-time indices $\mu,\nu, \cdots$ and $SO$(3,1) indices $a, b, \cdots$ run from 0 to 3. Time and space indices are indicated by $\mu=0, i$, and $a=(0), (i)$.}. Equation (\ref{q1}) leads to the metricity  condition and the identically vanishing of curvature tensor defined by ${\Gamma^\lambda}_{\mu\nu}$, given by equation  (\ref{q2}). The metric tensor $g_{\mu \nu}$ is defined by
\begin{equation}\label{q3}
 g_{\mu \nu} \stackrel{\textmd{def.}}{=}  \eta_{i j} {h^i}_\mu {h^j}_\nu,
\end{equation}
with $\eta_{i j}=(+1,-1,-1,-1)$ is the metric of Minkowski space-time. We note that, associated with any tetrad field ${h_i}^\mu$ there  is a metric field defined uniquely by (\ref{q3}), while a given metric $g^{\mu \nu}$ does not determine the tetrad field completely; for any local Lorentz transformation of the tetrads ${h_i}^\mu$ leads to a new set of tetrads which also satisfy (\ref{q3}). Defining the torsion components and the contortion as
\begin{eqnarray}
\nonumber {T^\alpha}_{\mu \nu}  & \stackrel {\textmd{def.}}{=} &
{\Gamma^\alpha}_{\nu \mu}-{\Gamma^\alpha}_{\mu \nu} ={h_a}^\alpha
\left(\partial_\mu{h^a}_\nu-\partial_\nu{h^a}_\mu\right),\\
{K^{\mu \nu}}_\alpha  & \stackrel {\textmd{def.}}{=} &
-\frac{1}{2}\left({T^{\mu \nu}}_\alpha-{T^{\nu
\mu}}_\alpha-{T_\alpha}^{\mu \nu}\right), \label{q4}
\end{eqnarray}
where the contortion equals the difference between Weitzen\"{o}ck  and Levi-Civita connection, i.e., ${K^{\mu}}_{\nu \rho}= {\Gamma^\mu}_{\nu \rho }-\left \{_{\nu  \rho}^\mu\right\}$.  The tensor ${S_\alpha}^{\mu \nu}$ is defined as
\begin{equation}\label{q5}
{S_\alpha}^{\mu \nu}
\stackrel {\textmd{def.}}{=} \frac{1}{2}\left({K^{\mu
\nu}}_\alpha+\delta^\mu_\alpha{T^{\beta
\nu}}_\beta-\delta^\nu_\alpha{T^{\beta \mu}}_\beta\right),
\end{equation}
which is skew symmetric in the last two indices. The torsion scalar is defined as
\begin{equation}\label{q6}
T \stackrel {\textmd{def.}}{=} {T^\alpha}_{\mu \nu}
{S_\alpha}^{\mu \nu}.
\end{equation}
Similar to the $f(R)$ theory, one can define the action of $f(T )$ theory as
\begin{equation}\label{q7}
{\cal L}({h^a}_\mu, \Phi_A)=\int
d^4x h\left[\frac{1}{16\pi}f(T)+{\cal L}_{Matter}(\Phi_A)\right], \quad \textrm
{where} \quad h=\sqrt{-g}=det\left({h^a}_\mu\right),
\end{equation}
and we assume the units in which $G = c = 1$  and  $\Phi_A$ are the matter fields.   Considering the action (\ref{q7}) as a function of the fields ${h^a}_\mu$ and putting  the variation of the function with respect to the field ${h^a}_\mu$ to be vanishing one can obtain the following equations of motion \cite{BF,CGSV}.
\begin{equation}\label{q8}
{S_\mu}^{\rho \nu} T_{,\rho} \
f(T)_{TT}+\left[h^{-1}{h^a}_\mu\partial_\rho\left(h{h_a}^\alpha
{S_\alpha}^{\rho \nu}\right)-{T^\alpha}_{\lambda \mu}{S_\alpha}^{\nu
\lambda}\right]f(T)_T-\frac{1}{4}\delta^\nu_\mu f(T)=-4\pi{{\cal T}^\nu}_\mu,
\end{equation}
where
$T_{,\rho}=\frac{\partial T}{\partial x^\rho}$, \; \; $f(T)_T=\frac{\partial f(T)}{\partial T}$, \; \;$f(T)_{TT}=\frac{\partial^2 f(T)}{\partial T^2}$ and ${{\cal T}^\nu}_\mu$ is the energy momentum tensor.

The relation between the Ricci scalar and the scalar torsion is given by \cite{DDS}--\cite{AM}
\begin{equation}\label{q9}
R=-T-2\nabla^\mu{T^\rho}_{\mu \rho}. \end{equation}
Equation (\ref{q9})  implies that the $T$ and $R$ differ only by a covariant divergence of a space time
vector. Therefore, the Einstein-Hilbert action and the teleparallel action (i.e. ${\cal L}_{TEGR} =
\int d^4x|e|T$) will
both lead to the same field equations and are dynamically equivalent theories. However, this term, divergence term, is the main reason that makes the field equations of $f(T)$ not invariant  under local Lorentz transformation. Let us explain this for some specific form of $f(T)$. If
\begin{eqnarray}\label{q10}   f(R)=&& R+R^2\equiv
 \left[-T-2\nabla^\mu{T^\rho}_{\mu \rho}\right]+\left[-T-2\nabla^\mu{T^\rho}_{\mu \rho}\right]^2\nonumber\\
 &&=-T-2\nabla^\mu{T^\rho}_{\mu \rho}+T^2+4\Biggl\{\left[\nabla^\mu{T^\rho}_{\mu \rho}\right]^2+T\nabla^\mu{T^\rho}_{\mu \rho}\Biggr\}.\end{eqnarray}
The last two terms in the R.H.S. of (\ref{q10})  is not a total derivative terms and therefore are responsible to make $f(R)=R+R^2$ when written in terms of $T$ and $T^2$ not invariant under local Lorentz transformation in contrast to the linear case, i.e., the form of (\ref{q9}) . Same discussion can be applied to the general form of $f(R)$ and $f(T)$ which shows in general a different between the $f(R)$ and $f(T)$ gravitational theories that makes the field equation of $f(R)$ to be of fourth order and invariant under local Lorentz transformation while $f(T)$ is of second order and not invariant under local Lorentz transformation.

\section{Cosmological Modifications of $f(T)$}\label{S3}

 In this study, we attempt  to apply the quadratic form of $f(T)$ field equations to the FRW universe. In this cosmological model the universe is taken as homogeneous and isotropic in space, which directly gives rise to the tetrad given by Robertson \cite{Rob}. This tetrad has the same metric as FRW metric; it can be written in spherical polar coordinate ($t$, $r$, $\theta$, $\phi$) as follows:
\begin{equation}\label{tetrad}
\left({h_{i}}^{\mu}\right)=\left(
  \begin{array}{cccc}
    1 & 0 & 0 & 0 \\
    0&\displaystyle\frac{L_1 \sin{\theta} \cos{\phi}}{4R(t)} & \displaystyle\displaystyle\frac{L_2 \cos{\theta} \cos{\phi}-4r\sqrt{k}\sin{\phi}}{4 r R(t)} & -\displaystyle\frac{L_2 \sin{\phi}+4 r \sqrt{k} \cos{\theta} \cos{\phi}}{4 r R(t)\sin{\theta}} \\[5pt]
    0&\displaystyle\frac{L_1 \sin{\theta} \sin{\phi}}{4 R(t)} & \displaystyle\frac{L_2 \cos{\theta} \sin{\phi}+4 r \sqrt{k}\cos{\phi}}{4 r R(t)} & \displaystyle\frac{L_2 \cos{\phi}-4 r \sqrt{k} \cos{\theta} \sin{\phi}}{4 r R(t)\sin{\theta}} \\[5pt]
    0&\displaystyle\frac{L_1 \cos{\theta}}{4 R(t)} & \displaystyle\frac{-L_2 \sin{\theta}}{4 r R(t)} & \displaystyle\frac{\sqrt{k}}{R(t)} \\[5pt]
  \end{array}
\right),
\end{equation}
where $R(t)$ is the \textit{scale factor}, $L_1=4+k r^{2}$ and $L_2=4-k r^{2}$\footnote{Tetrad (12) is a combination of the diagonal tetrad and so(3), i.e.\begin{eqnarray} & & \left( {h_i}^\mu \right)=
\left({\Lambda_i}^j\right)\left(
{h_j}^\mu \right)_1 \qquad {\textrm where} \nonumber\\
& & \left({\Lambda_i}^j\right)=\left(  \begin{array}{cccc} 1 &  0 & 0 & 0
\\[5pt]  0  &  \sin\theta \cos\phi &  \cos\theta \cos\phi &
- \sin\phi \\[5pt] 0  & \sin \theta \sin \phi& \cos\theta
\sin\phi & \cos\phi \\[5pt] 0  & \cos\theta & -\sin\theta &
0  \end{array}\right)\;,  \nonumber\\
& &   \left( {h_j}^\mu \right)_1=\left(  \begin{array}{cccc}
 1& 0 & 0 & 0 \\[5pt] 0
&  \frac{1}{R(t)} &0& 0
\\[5pt] 0  & 0&\frac{1}{rR(t)} &0\\[5pt] 0  & 0 & 0 &
\frac{1}{R(t)r\sin\theta}  \end{array}\right)\;. \nonumber \end{eqnarray}  }. Substituting from the vierbein (\ref{tetrad}) into (\ref{q6}), we get the torsion scalar
\begin{equation}\label{Tscalar}
\begin{split}
   T=&\frac{6 k- 6 \dot{R}^2}{R^2},\\
    =&-6\left(H^2-\frac{k}{R^2},\right)\\
    =&-6H^2(1+\Omega_{k}),
\end{split}
\end{equation}
where $H(=\frac{\dot{R}}{R})$ is the \textit{Hubble} parameter and $\Omega_{k}(=\frac{-k}{R^2 H^2})$ is the \textit{curvature} energy density parameter. This form of $T$, Eq. (\ref{Tscalar}),  would be able to give a proper reduced Lagrangian for the dynamics of the scale factor $R(t)$, as a
consequence of its independence on  the radial coordinate $r$. This $r$-independent Weitzenb\"ock invariant is  consistent with the isotropy and
homogeneity of the FRW cosmological models. The field equations (\ref{q8}) read
\begin{equation}\label{T00}
    \mathcal{T}_{0}^{~0}=\frac{-R^{2} f -12\dot{R}^{2} f_{T}}{4 R^{2}},
\end{equation}
\begin{equation}\label{T11}
\mathcal{T}_{1}^{~1}=\mathcal{T}_{2}^{~2}=\mathcal{T}_{3}^{~3}=\frac{4 k (R^{2} f_{T}+12 \dot{R}^{2} f_{TT})-R^{4} f -4R^2(R\ddot{R}+2\dot{R}^2)f_{T}+48\dot{R}^2(R\ddot{R}-\dot{R}^2)f_{TT}}{4 R^4},
\end{equation}
where the EoS is taken for a perfect fluid so that the energy-momentum tensor is ${\mathcal{T}^{\mu}}_{\nu}=\textmd{diag}(\rho,-p,-p,-p)$. Using (\ref{T00}), the perfect fluid density $\rho$ is given by
\begin{equation}\label{dens1}
    4 \pi \rho=\frac{R^{2} f +12\dot{R}^{2} f_{T}}{4 R^{2}},
\end{equation}
and using (\ref{T11}), the proper pressure $p$ of the perfect fluid is given by
\begin{equation}\label{press1}
    4 \pi p=\frac{4 k (R^{2} f_{T}+12 \dot{R}^{2} f_{TT})-R^{4} f -4R^2(R\ddot{R}+2\dot{R}^2)f_{T}+48\dot{R}^2(R\ddot{R}-\dot{R}^2)f_{TT}}{4 R^4}.
\end{equation}
Equations (\ref{dens1}) and (\ref{press1}) are the modified Friedmann equations in the $f(T)$-gravity in its generalized form.
\subsection{The flat FRW dynamical equations}\label{S3.1}

Let us assume  that the background is a non-viscous fluid and also assume the flat case, i.e., $k=0$. Then from (\ref{dens1}) and (\ref{press1}) we get
\begin{equation}\label{densf1}
    4 \pi \rho=\frac{R^{2} f +12\dot{R}^{2} f_{T}}{4 R^{2}},
\end{equation}
and using (\ref{T11}), the proper pressure $p$ of the perfect fluid is given by
\begin{equation}\label{pressf1}
    4 \pi p=\frac{48\dot{R}^2(R\ddot{R}-\dot{R}^2)f_{TT}-4R^2(R\ddot{R}+2\dot{R}^2)f_{T}-R^{4} f}{4 R^4}.
\end{equation}
Alternatively, we can study the torsion contribution to both $\rho$ and $p$ in the Friedmann dynamical equations by replacing $\rho \rightarrow \rho+\rho_{T}$ and $p \rightarrow p+p_{T}$, where $\rho$, $\rho_{T}$,  $p$ and $p_{T}$ are the matter density, the torsion density, the matter pressure and the torsion pressure respectively.
\begin{equation}\label{FRW1}
  3\left(\frac{\dot{R}}{R}\right)^2=3H^2  =8\pi \rho + 8\pi \rho_{T}, \end{equation}
  \begin{equation}\label{FRW2}
  3\left(\frac{\ddot{R}}{R}\right) =3 q H^2 = -4 \pi \left(\rho+3 p\right)-4 \pi \left(\rho_{T}+3 p_{T}\right),
\end{equation}
where $q(=-\frac{R\ddot{R}}{\dot{R}^2})$ is the \textit{deceleration} parameter. In the above equation we take the general case of a non-vanishing pressure $p \neq 0$. It is clear that when $\rho_{T}=0$ and $p_{T}=0$ the above equations reduce to the usual Friedmann equations in GR. We take $\rho=\rho_{c}$ where $\rho_{c}$ is the critical density of the universe when it is full of matter. Substituting in equations (\ref{FRW1}) and (\ref{FRW2}) we get
\begin{eqnarray}
  1 &=& \Omega_{m}+\Omega_{T},\label{FRW3} \\
  q &=& \frac{\left(\rho+3 p\right)/2}{3H^2/8 \pi}+\frac{\left(\rho_{T}+3 p_{T}\right)/2}{3H^2/8 \pi}, \label{FRW4}
\end{eqnarray}
where $\Omega_{m}=\frac{\rho}{\rho_{c}}=\frac{\rho}{3H^2/8\pi}$ represents the \textit{matter density} parameter and $\Omega_{T}=\frac{\rho_{T}}{\rho_{c}}=\frac{\rho_{T}}{3H^2/8\pi}$ represents the \textit{torsion density} parameter.

In order to obtain the torsion contribution $\rho_{T}$ and $p_{T}$, we rewrite equations (\ref{densf1}) and (\ref{pressf1}), in terms of the Hubble parameter, as below
\begin{equation}\label{dens2}
    4 \pi \rho=\frac{1}{4}(f+12 H^2 f_{T}).
\end{equation}
\begin{equation}\label{press2}
    4 \pi p=12 \dot{H} H^2 f_{TT}-\left(\dot{H}+3H^2\right)f_{T}-\frac{1}{4}f.
\end{equation}
Substituting the matter density that is obtained by the $f(T)$ field equation (\ref{dens2}) into the FRW dynamical equation (\ref{FRW1}), we get the torsion density
\begin{equation}\label{Tor_density}
    \rho_{T}=\frac{1}{8 \pi}\left(3H^2-f/2-6H^2 f_{T}+\frac{3k}{R^2}\right).
\end{equation}
The above equation can be written in the form
\begin{equation*}
    \frac{\rho_{T}}{3H^2/8\pi}=1-\left[\frac{f}{6 H^2}+2 f_{T}\right]+\frac{k}{H^2 R^2},
\end{equation*}
so that the torsion density parameter is
\begin{equation}\label{tor_dens_par}
    \Omega_{T}=1-\left[\frac{f}{6 H^2}+2 f_{T}\right],
\end{equation}
comparing the above equation to equation (\ref{FRW3}) we get the modified matter density parameter as
\begin{equation}\label{matt_density_par}
    \Omega_{m}=\frac{f}{6H^2}+2 f_{T}.
\end{equation}
Similarly we substitute from (\ref{dens2}), (\ref{press2}) and (\ref{Tor_density}) into (\ref{FRW2}) we get
\begin{equation}\label{Tor_press}
    p_{T}=\frac{-1}{8 \pi}\left[2\dot{H}+3H^2-f/2-2(\dot{H}+3H^2)f_{T}+24\dot{H}H^2 f_{TT}\right].
\end{equation}
The EoS parameter due to the torsion contribution is thus
\begin{equation}\label{Tor_EoS_par}
    \omega_{T}=\frac{p_{T}}{\rho_{T}}=-1+2/3\frac{(1-f_{T}+12H^2f_{TT})\dot{H}}{f/6-(1+2f_{T})H^2}.
\end{equation}
It is clear that $\omega_{T}=-1$  when $\dot{H}=0$ \cite{KA12}. The torsion contributes to the FRW model in a way similar to the cosmological constant.

When $f(T)$ takes the form
\begin{equation}\label{f(T_)}
f(T)=T+\epsilon T^2,
\end{equation}
equations (\ref{dens1}) and  (\ref{press1}) take the form
\begin{eqnarray}\label{press3}
&&    4 \pi \rho=\frac{1}{4}[T+\epsilon T^2+12 H^2(1+2\epsilon T)], \label{dpq}\nonumber \\
  && 4 \pi p =24 \dot{H} H^2 \epsilon-\left(\dot{H}+3H^2\right)(1+2\epsilon T)-\frac{1}{4}[T+\epsilon T^2].
  \end{eqnarray}
Using the equation of state
\begin{equation}\label{es}
p=\omega \rho, \end{equation} where $\omega=-1, \; 0,\; 1/3,\; 1$ for dark energy, dust, radiation and stiff matter respectively.
\subsection{Dark energy case}\label{S3.2}

 Using (\ref{dpq}) in (\ref{es}) we get for the $\omega=-1$
\begin{equation}\label{est}
24 \dot{H} H^2 \epsilon-\dot{H}(1+2\epsilon T)=0. \end{equation}
The solution of the above differential equation has many possible cases: We exclude the case of $R(t)$ is a constant as it gives a steady universe.  By solving for the scale factor $R(t)$ we get
\begin{equation}\label{so}
R(t)=e^\frac{c-t}{6\sqrt{\epsilon}},\end{equation}
where $c$ is a constant of integration. Consequently, we can evaluate
\begin{equation}\label{cos}
T:=\frac{-1}{6\epsilon}, \qquad H:=\frac{-1}{6\sqrt{\epsilon}}, \qquad q:=-1, \qquad \omega_T=\frac{p_T}{\rho_T}=-1. \end{equation}
The solution gives an exponential expanding universe with a constant Hubble parameter, i.e. de Sitter universe\footnote{It is not allowed to put the parameter $\epsilon=0$ in solution (\ref{so}). This is because that when $\epsilon=0$,  Eq.  (\ref{so}) gives zero scale factor when $c<t$ and infinite scale factor when $c>t$. Therefore,   the parameter $\epsilon$ has a finite value. }. So equivalently we have a vacuum dominant universe which powered the inflation scenario. Recalling (\ref{f(T_)}), we find the coefficient $\epsilon$ of the $T^{2}$ higher order gravity contributes to this inflationary phase. The comparison of the results of (\ref{cos}) with the de Sitter solution directly relates the coefficient $\epsilon$ to the cosmological constant $\Lambda$ by $\epsilon = \frac{1}{12 \Lambda}$. It is well known that at this early stage the cosmological constant has a value much larger than its present value. The discrepancy between the two values is about 120 orders of magnitude, this is called the cosmological constant problem. According to our analysis here; we expect the value of $\alpha$ is very small $\sim 10^{-73}~s^{2}$, so that the quadratic contribution in the $f(T)$ can be ignored.

This case shows that both the matter and the torsion fluids having the same behavior where $\rho=\rho_{T}=\frac{1}{192\pi \epsilon}=\frac{\Lambda}{16\pi}$, which is the density of vacuum. Also, the EoS parameters has been obtained as $\omega=\omega_{T}=-1$. This case gives directly a de Sitter universe, where both matter and torsion act as a vacuum state in an indistinguishable behavior. Using  (\ref{f(T_)}) and (\ref{cos}) we get $f(T)=-\frac{5}{36\epsilon}\sim T$. Interestingly, we find that the solution omits the second order contribution of the $f(T)$ naturally and the $f(T)$ is a constant, more precisely it acts just like the cosmological constant. This may explain why the two-fluid components have the same behavior. We saw that the choice of $\omega=-1$ of the matter fluid constrains the torsion fluid to acquire the same behavior producing a pure vacuum universe which represents the key of the inflationary universe. Although the model does not address the cosmological constant issue, it succeeded to predict the exponential expansion of the inflation models. But it does not provide a mechanism to end the inflation period and resuming the big bang to radiation dominant universe era.

\subsection{Matter dominate  case}\label{S3.1}
In this case we are going to solve  (\ref{es}) using (\ref{press3}) for the general case $\omega\neq -1$. The exact solution has the form
\begin{equation}\label{sc_fac_matt}
    R(t)=c_1\underbrace{\Biggl[(1+\omega)(t-t_0)-6\epsilon\Biggr]^{\frac{2}{3(1+\omega)}}}_{matter}=c_1R_m,
\end{equation}
where $c_{1}$  is arbitrary constant of integration and $t_{0}$ represent the present cosmic time, with an initial condition
$H_0 = H(t_0)$. The scale factor  (\ref{sc_fac_matt}),  represents the matter dominant epochs, it takes the values of $\Biggl[4(t-t_0)/3-6\epsilon\Biggr]^{\frac{1}{2}}$  and $\Biggl[(t-t_0)-6\epsilon\Biggr]^{\frac{2}{3}}$ for radiation ($\omega=1/3$) and dust ($\omega=0$) states, respectively. This values asymptotically approach $t^{1/2}$ and $t^{2/3}$.
 \begin{figure}
\begin{center}
\includegraphics[scale=.3]{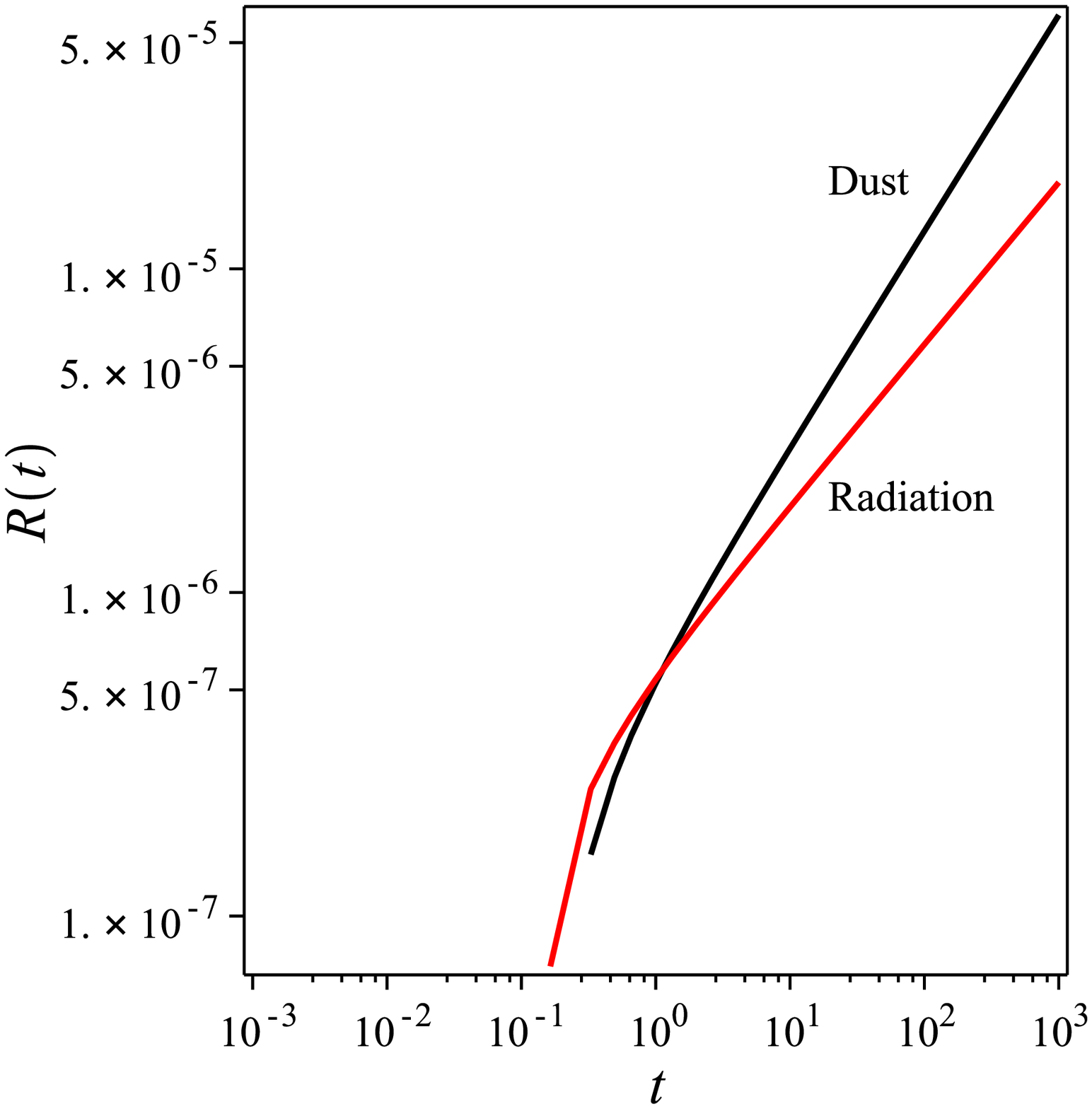}
\caption{ Phase transitions in the late universe.  The red line shows the behavior of $R_{m}$ when the matter has been chosen as a radiation , i.e. $\omega=1/3$. The black line shows the behavior of $R_{m}$ when the matter has been chosen as a dust , i.e. $\omega=0$.  The initial values have been chosen as $c_{1}\sim 10^{-44}$, $c_{2}\sim 10^{-7}$. The model parameter is chosen as $\epsilon=10^{-2}$.}
\label{Fig1}
\end{center}
\begin{center}
\includegraphics[scale=.3]{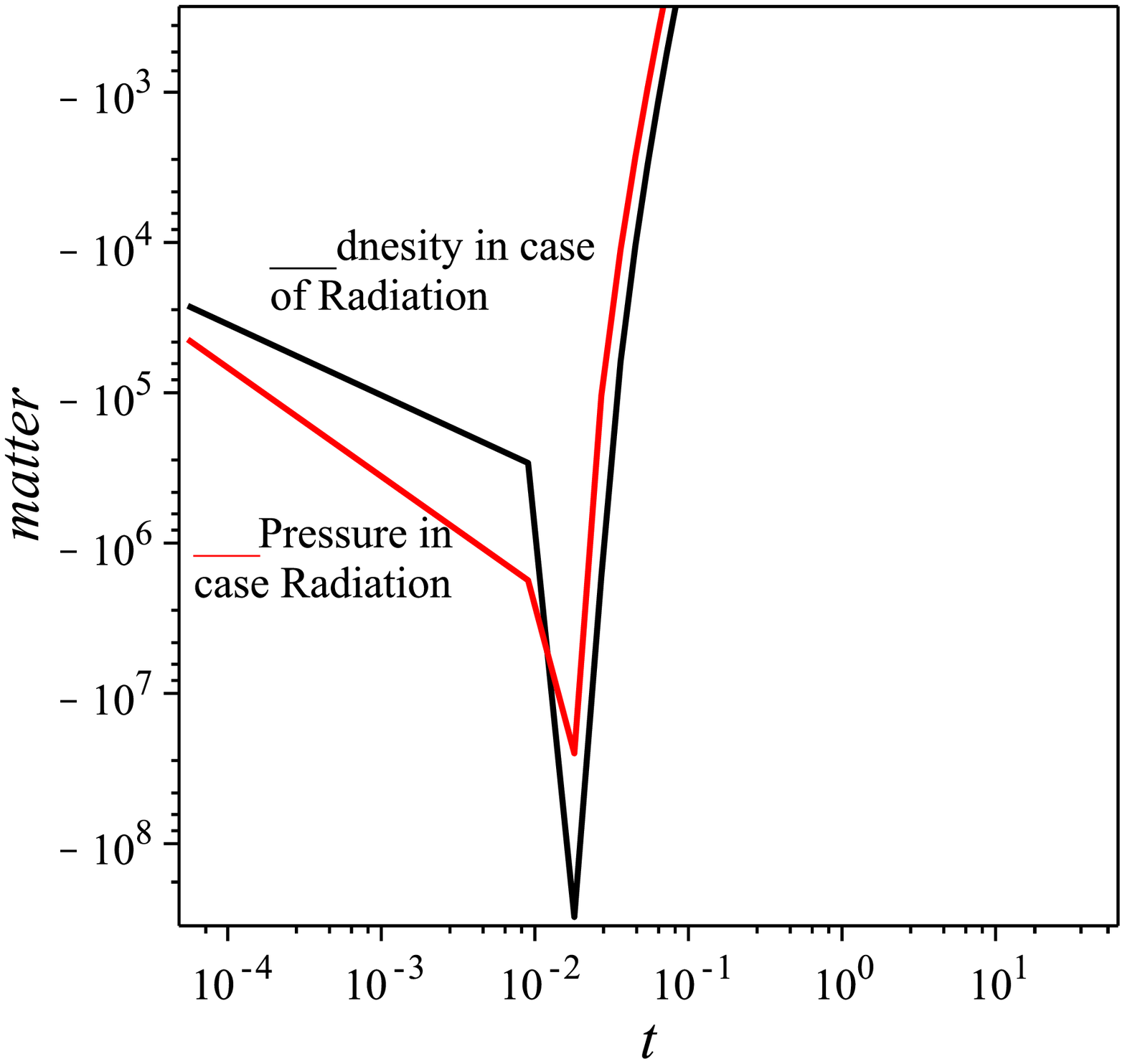}
\caption{ The evolution of density and pressure for radiation case. The initial values have been chosen as $c_{1}\sim 10^{-44}$, $c_{2}\sim 10^{-7}$. The model parameter is chosen as $\epsilon=10^{-2}$.}
\end{center}
\end{figure}
The above limits show that the scale factor is dominated by matter  at the late  universe
time. This is consistent with the late behavior of the  scale factor that matches perfectly the radiation
dominant, till the recombination at the radiation-dust equality (i.e. $R_{radiation}= R_{dust}$) where the universe is
cooled enough to turn on the non-relativistic (dust) matter.
The torsion density (\ref{Tor_density}) and pressure (\ref{Tor_press}) read
\begin{equation}\label{Tor_densk+1}
    \rho_T=\frac{4\epsilon}{3 \pi([1+\omega][t-t_0]-2\alpha)^4}, \qquad p_T=\frac{4\epsilon(1+2\omega)}{3 \pi([1+\omega][t-t_0]-2\alpha)}.
\end{equation}
\section{Cosmological parameters}\label{S4}
The standard cosmology gives a set of parameters to study the evolution of the universe through the Hubble and the
deceleration parameters, while the other parameters are called density parameters allow studying the composition
of the universe. We study some of these parameters correspond to the  scale factor (\ref{sc_fac_matt}) and the assumed
two-fluids which are given by the set of equations (\ref{FRW1}) and (\ref{FRW2}). So the Hubble parameter reads
\begin{equation}
H(t)=\frac{2}{3(1+\omega)(t-t_0)-6\epsilon}.\end{equation}
One can see that when the initial value $t_0$ is chosen as the Planck time, the Hubble parameter counts a large value
at this late time, while it decays at early time, $limt_{t=\infty}=0$. Interestingly, the above expression shows a
time dependent Hubble parameter capable to describe  later matter dominant phases. Also,
the deceleration parameter shows a consistent results with the above description. We evaluate the deceleration
parameter as
\begin{equation} q=\frac{1}{2}+\frac{3\omega}{2}.\end{equation}
 The
asymptotic behavior is consistent with the known results
as $q\rightarrow\frac{1}{2}$ or 1 when the EoS of the matter are chosen as
$(\omega = 0)$ for dust or $(\omega = 1/3)$ for radiation, respectively.
We evaluate the teleparallel torsion scalar
as
\begin{equation}
T(t)=\frac{-24}{[3(1+\omega)(t-t_0)-6\epsilon]^2}.\end{equation}
We
study when the matter is dust and radiation so  the torsion EoS reads
\begin{equation} \omega_T=1+2\omega,\end{equation}
where the asymptotic
behavior of the torsion EoS approaches the values of
$\omega_T\rightarrow 1$ or $\omega_T\rightarrow 5/3$, respectively.

 Finally, we study the effective EoS $\omega_{\textmd{eff}}:=\frac{p+p_{T}}{\rho+\rho_{T}}$. Using the equations (\ref{dens2})-(\ref{Tor_density}) and (\ref{Tor_press}) we obtain
\begin{eqnarray}
\omega_{\textmd{eff}}:=\omega.
\end{eqnarray}
The above limits show that the effective EoS  turns itself to matter dominant phase with no need to slow-roll conditions.
\begin{figure}
\begin{center}
\includegraphics[scale=.4]{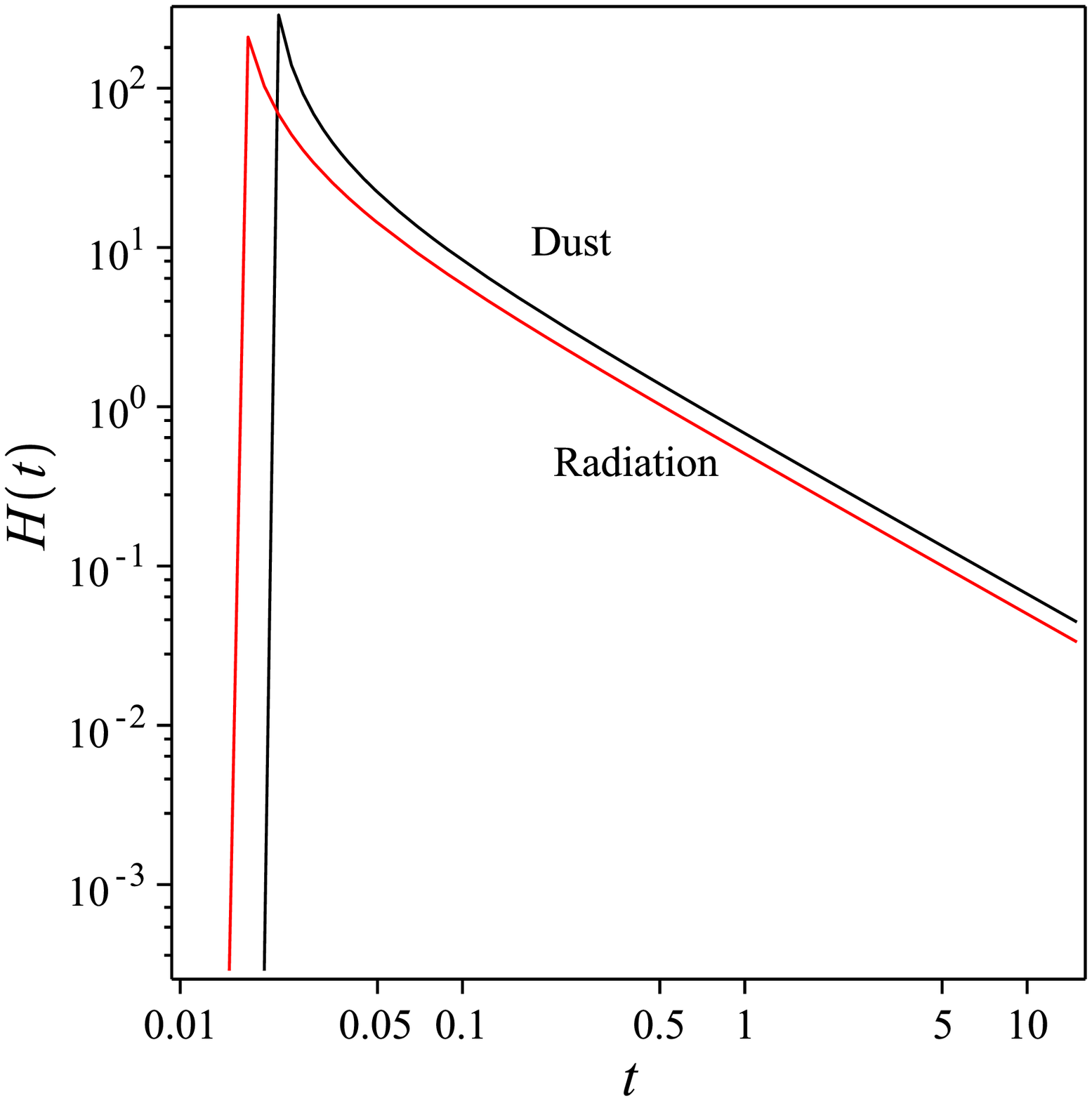}
\caption{ The evolution of the Hubble parameter (4.1) indicates an early cosmic inflation.}
\label{Fig2}
\end{center}
\begin{center}
\includegraphics[scale=.4]{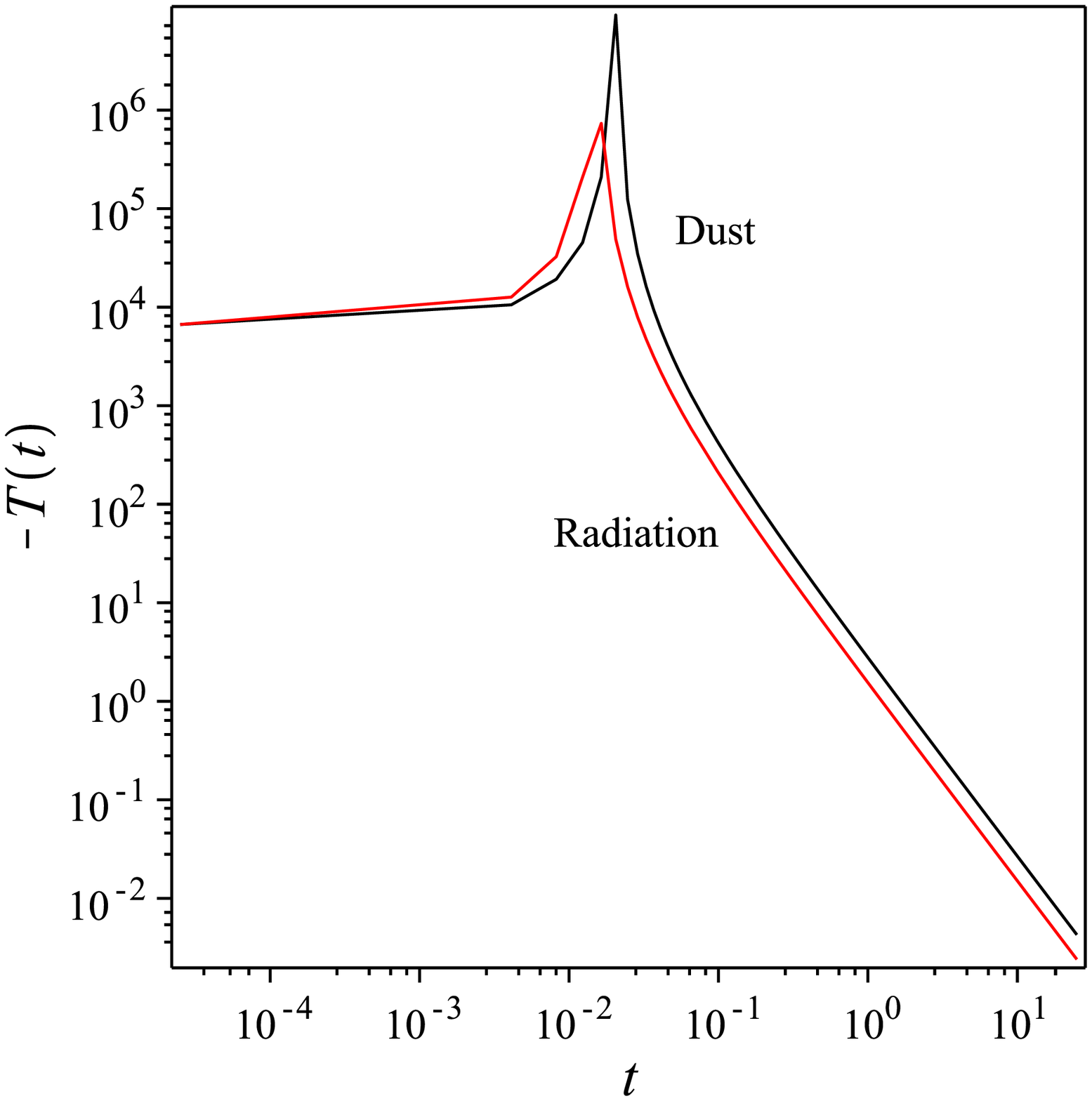}
\caption{ the decay of the teleparallel torsion scalar field (4.3).}
\label{Fig3}
\end{center}
\end{figure}
\section{Concluding Remarks}\label{S5}
 We have used a quadratic form of $f(T)=T +\epsilon T^2$ field equations to the flat
FRW universe, i.e., the sectional curvature $k=0$, with a perfect fluid. This choice enabled us to amended  the FWR differential equations. This adaptation is due to the torsion contribution 
where  the density and pressure of two fluids of matter and
torsion reproduce   TEGR as the parameter $\epsilon\rightarrow 0$. We  assumed the
matter to be controlled by the equation of state. We have studied
two cases of the matter choices:\vspace{.2cm}\\
      i) The universe is assumed to be dominant by $\Lambda$DE with EoS $\omega=-1$. The solution provided an exponential scale factor $R(t)=e^\frac{c-t}{6\sqrt{\epsilon}}$, which fixes the torsion EoS to a value of $\omega_{T}=-1$. The standard cosmological study has given Hubble and deceleration parameters as $H=const.$ and $q=-1$. So this model is consistent with $\Lambda$ de Sitter model. Also, we found that the choice of $\omega=-1$ implying that $f(T) \sim T \sim \Lambda$. Consequently, the coefficient $\epsilon$ has been estimated as $\epsilon=\frac{1}{12 \Lambda} \sim 10^{-73}~s^{2}$. So the model omitted the quadratic term $T^{2}$ anyways. We found that the model can predict an eternal inflationary scenario. From this case we can conclude that the higher order of torsion acts as a cosmological constant and this is consistent with literature \cite{CDDS1,CDDS}  \vspace{.2cm}\\
     ii) The universe is assumed to have an EoS $\omega \neq -1$. The solution provided a scale factor,  whereas $R \propto t^{\frac{2}{3(1+\omega)}}$ matches perfectly the matter dominant universe epoch.  The  scale factor matches perfectly the matter scale factor announcing a matter dominant universe to showup.

\subsection*{ACKNOWLEDGMENTS}
This work is partially supported by the Egyptian Ministry of Scientific Research under project No. 24-2-12.

\end{document}